\documentclass[apj]{emulateapj}
\usepackage{xspace,graphics,graphicx}
\usepackage{lscape}
\usepackage{natbib}

\def\sci#1{{\; \times \; 10^{#1}}}
\newcommand{\lya}{Ly$\alpha$}
\newcommand{\mkms}{{\rm \; km\;s^{-1}}}

\slugcomment{Submitted to ApJ} 

\shorttitle{Galaxies probing galaxies}
\shortauthors{Rubin et al.}

\begin{document}

\title{Galaxies probing galaxies: cool halo gas from a $z = 0.47$ post-starburst galaxy\altaffilmark{1}}
\author{Kate H. R. Rubin\altaffilmark{2,3}, J. Xavier Prochaska\altaffilmark{2}, David C. Koo\altaffilmark{2}, Andrew C. Phillips\altaffilmark{2} \& Benjamin J. Weiner\altaffilmark{4}} 

\altaffiltext{1}{Some of the data presented herein were obtained at the W. M. Keck Observatory, which is operated as a scientific partnership among the California Institute of Technology, the University of California and the National Aeronautics and Space Administration.  The Observatory was made possible by the generous financial support of the W.M. Keck Foundation.}
\altaffiltext{2}{University of California Observatories, University of California, Santa Cruz, CA 95064}  
\altaffiltext{3}{rubin@ucolick.org}
\altaffiltext{4}{Steward Observatory, 933 N. Cherry St., University of Arizona, Tucson, AZ 85721}

\begin{abstract}

We study the cool gas around a galaxy at $z = 0.4729$ using Keck/LRIS spectroscopy of a bright ($B = 21.7$) background galaxy at $z = 0.6942$ at a transverse distance of $16.5~ h_{70}^{-1}$ kpc. 
The background galaxy spectrum reveals strong \ion{Fe}{2}, \ion{Mg}{2}, \ion{Mg}{1}, and \ion{Ca}{2} absorption at the redshift of the foreground galaxy, with a \ion{Mg}{2} $\lambda 2796$ rest equivalent width 
of $3.93 \pm 0.08$ \AA, indicative of a velocity width exceeding $400~\rm km~s^{-1}$.  Because the background galaxy is large ($ > 4~ h_{70}^{-1}$ kpc), the high covering fraction of the absorbing gas suggests that it arises in a spatially extended complex of cool clouds with large velocity dispersion.  Spectroscopy of the massive ($\rm \log M_*/M_{\odot} = 11.15 \pm 0.08$) host galaxy reveals that it experienced a burst of star formation about 1 Gyr ago and that it harbors a weak AGN.  We discuss the possible origins of the 
cool gas in its halo, including multiphase cooling of hot halo gas, cold inflow, tidal interactions, and galactic winds.  We conclude the absorbing gas was most likely ejected or tidally stripped from the interstellar medium of the host galaxy or its progenitors during the past starburst event.  Adopting the latter interpretation, these results place one of only a few constraints on 
the radial extent of cool gas driven or stripped from a galaxy 
in the distant Universe.
Future studies with integral field unit spectroscopy of spatially extended background galaxies will provide multiple sightlines through foreground absorbers and permit analysis of the morphology and kinematics of the gas surrounding galaxies with a diverse set of properties and environments.

  \end{abstract}
\keywords{galaxies: absorption lines --- galaxies: evolution --- galaxies: halos}

\section{INTRODUCTION}

Several decades ago, \citet{BahcallSpitzer1969} postulated 
that the absorption lines observed in the spectra of 
distant QSOs are due to the extended gaseous halos of 
intervening galaxies.  Since that time, astronomers have 
identified the galaxies associated with the observed absorption \citep[e.g.,][]{Yanny1990,BergeronBoisse1991,Steidel1995} and
have used QSO spectroscopy of various rest-frame ultraviolet transitions to  
constrain the
nature of baryonic processes in the outer regions of galaxies,
e.g., feedback, accretion, and cooling \citep{Mo1996,MB2004,Tobias2006}.
These transitions include \lya\ and \ion{Mg}{2} lines that probe cooler,
photoionized gas \citep{Wolfe1995,Bergeron1986}, and \ion{C}{4} and \ion{O}{6}
doublets that trace warmer, more diffuse material 
\citep{Chen2001,Tripp2008}. 
However, while the apparent brightness of QSOs 
enables high
signal-to-noise, high spectral resolution datasets,
they severely outshine the galaxies projected nearby and therefore limit
follow-up analysis, especially at small impact parameters.
Furthermore, QSOs  
are too rare in the sky
to probe numerous individual galaxies with
multiple sightlines.  Therefore,  
constraints on the
covering fraction ($C_f$) and spatial distribution of halo gas 
must be statistical \citep[e.g.,][]{Chen2001,CT2008}.

The \ion{Mg}{2} $\lambda \lambda 2796, 2803$ doublet in particular has been studied extensively, as it is easily accessible at optical wavelengths at redshifts $0.3 \lesssim z \lesssim 2.3$ and traces cool gas \citep[with temperature $\rm T \sim 10^4~K$;][]{BergeronStasinska1986} in a broad range of neutral hydrogen column densities \citep[$N($\ion{H}{1}$) = 10^{18} - 10^{22}~\rm cm^{-2}$;][]{Churchill2000,Rao2006}.  In spite of this, the physical origins of the gas giving rise to \ion{Mg}{2} absorption remain obscure.
Several studies \citep[e.g.,][]{MB2004} suggest an infall origin for this gas; for example, \citet{TC2008} present an infall model with which they reproduce the observed frequency distribution function for \ion{Mg}{2} systems as well as their clustering properties \citep{Bouche2006,Gauthier2009}.  
Other studies \citep{Bond2001,Schaye2001,Bouche2006,Bouche2007,MenardChelouche2009} 
suggest instead that these systems arise in gas which has been blown out of star-forming galaxies via superwinds.
In support of this latter scenario, a recent study 
by \citet{Weiner2009} of \ion{Mg}{2} kinematics in coadded DEEP2 spectra of a sample of star-forming galaxies at $z \sim 1.4$  reveals frequent, perhaps even ubiquitous, outflows of cool gas.   
\citet{Tremonti2007} also detect very high velocity outflows in $z \sim 0.6$ post-starburst galaxies traced by \ion{Mg}{2} absorption.  Such winds may redistribute \ion{Mg}{2} absorbing gas, possibly to a galaxy's halo, although the distances to which the winds extend remain uncertain.  
In one of the only studies addressing this issue, \citet{Martin2006} observes cool outflowing gas 
several kiloparsecs from the nuclei of a sample of $z \sim 0.1$ ultraluminous infrared galaxies. 

In principle, galaxy spectra can also probe the halo gas of foreground galaxies along the sightline \citep[e.g.,][]{Adelberger2005,Barger2008}.  This novel technique offers several advantages over QSO-galaxy pair studies (D. Koo et al. 2009, in preparation).  The projected number density of galaxies on the sky is much greater than that of QSOs; therefore the use of galaxies can vastly increase the number of potential background probes for a given galaxy halo.  Moreover, many galaxies are extended sources which provide the opportunity to study gas along multiple lines of sight through a given foreground halo, including their own \citep[e.g.,][]{Phillips1993,Heckman2000}.  Integral Field Unit (IFU) spectrographs may then be used to study the morphology of halo absorption and the spatial distribution of outflowing gas.   

Close transverse pairs of galaxies may be identified in large spectroscopic or photometric surveys so that a targeted search for foreground \ion{Mg}{2} absorption in the background galaxy spectra can be performed.  
This technique has been used at $z > 1.5$ to study absorbing gas traced by transitions such as \ion{C}{4} near Lyman Break Galaxies 
 by \citet{Adelberger2005}; but it has not yet been used at lower redshifts where higher resolution imaging and spectroscopy can be obtained for each galaxy.
 
In the process of carrying out a spectroscopic survey to measure outflow properties in galaxies at $0.3 < z < 1.4$, we identified a close transverse pair of galaxies in which the spectroscopy of the more distant object ($z = 0.6942$) 
allows the detection (in absorption) of gas in the environs of a foreground 
galaxy (at $z = 0.4729$ and at impact parameter $\rho = 16.5~ h_{70}^{-1} $~kpc). 
This foreground absorber shows signs of recent merger activity, has a stellar
continuum consistent with that of a post-starburst galaxy, and is host
to a low-luminosity AGN.
Here we analyze the spectrum of the background galaxy to examine halo gas in the foreground system.
We discuss our observations of the galaxy pair and data reduction in \S\ref{sec.thepair}.  Analysis of the luminous components of the galaxies and the foreground halo absorption is given in \S\ref{sec.analysis}.  The possible origins of the observed cool halo gas are discussed in \S\ref{sec.discussion}, and we conclude in \S\ref{sec.conclusions}.
We adopt a $\rm \Lambda CDM$ cosmology with $h_{70} = H_0 / 70~\rm km~s^{-1}~Mpc^{-1} $, $\rm \Omega_M = 0.3$, and $\rm \Omega_{\Lambda} = 0.7$.  Where it is not explicitly written, we assume $h_{70} = 1$.  Magnitudes quoted are in the AB system.

\section{THE GALAXY PAIR}\label{sec.thepair}

Our targeted galaxy pair is located in the GOODS-N field \citep[Great Observatories Origins Survey;][]{Giavalisco2004} and has been imaged by the HST Advanced Camera for Surveys in 4 optical bands (F435W, F606W, F775W and F850LP, or $B_{435}$, $V_{606}$, $i_{775}$ and $z_{850}$).  Galaxy properties derived in previous studies and references are given in Table~\ref{tab.photinfo}.

\begin{figure*}[ht]
\begin{center}
\includegraphics[width=6in]{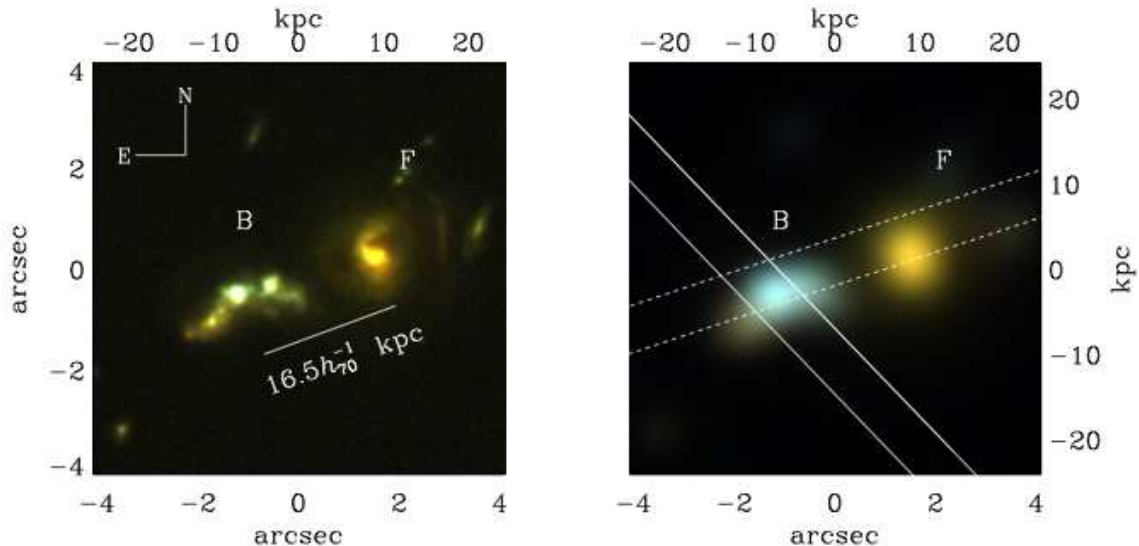}
\caption{{\it (Left)} Color image of the galaxy pair in the HST/ACS
$B_{435}$, $V_{606}$ and $i_{775}$ bands.  TKRS4389 is to the southeast in the image and marked with a ``B"; TKRS4259 is to the northwest and marked with an ``F".  These galaxies are offset by an impact parameter of 2.8\arcsec; i.e., $16.5~h_{70}^{-1}~ \rm kpc$ at the foreground redshift.  {\it (Right)}  
Same images as on the left, convolved with a $\rm FWHM = 0.8\arcsec$ Gaussian to simulate the effects of seeing.  The orientations of the slitlet used to obtain the $2.9$ hour spectrum (position angle $\sim 45^{\circ}$ East of North) and the longslit used to obtain the 33 minute spectrum (position angle $\sim 110^{\circ}$ East of North) are shown in white solid and dashed lines, respectively.  The position of the slitlet through B was obtained from the TKRS astrometry.  The position of the longslit was inferred by assuming the position angle given in the LRIS image header, convolving the HST $V_{606}$-band image with the seeing disk, and adjusting the slit position on the convolved image until the spatial profile best matched the profiles actually observed in the LRIS longslit between 5500 and 7100 \AA.
\label{fig.im}}
\end{center}
\end{figure*}

We obtained spectroscopy of both galaxies using the Low Resolution Imaging Spectrometer (LRIS) on Keck 1 \citep{Cohen1994} on 2008 May 30-31 UT.  We used the $\rm 600~ l~mm^{-1}$ grism blazed at 4000~\AA~on the blue side and the $\rm 600~ l~mm^{-1}$ grating blazed at 7500~\AA~on the red side with the D560 dichroic.  This setup affords a FWHM resolution ranging between $400~\rm km~s^{-1}$  and $180~\rm km~s^{-1}$ and wavelength coverage between $\sim 3200$~\AA~ and $\sim 7600-8000$~\AA.  Two sets of spectra were obtained.  We first used a slitmask to place a $0.9\arcsec$ slitlet across the center of the background galaxy oriented NE and collected $6 \times \sim 1800\rm~ sec$ exposures with FWHM $\sim 0.6\arcsec$ seeing.  We also placed a $1\arcsec$-wide longslit across both galaxies and obtained $2 \times 1000\rm~ sec$ exposures with $\sim 1\arcsec$ seeing (see Figure~\ref{fig.im}). 

The data were reduced using the XIDL LowRedux\footnote{http://www.ucolick.org/$\sim$xavier/LowRedux/} data reduction pipeline.  The pipeline includes bias subtraction and flat-fielding, slit finding, 
wavelength calibration, object identification, sky subtraction, cosmic ray rejection, and flux calibration.  Vacuum and heliocentric corrections were applied.  Because the separation between the galaxies is only 2.8\arcsec, we used a spatially narrow Gaussian centered on the foreground object to weight the extracted spectrum (with $\rm FWHM \sim 0.54\arcsec$ and $0.75\arcsec$ for the blue and red sides, respectively).  This limits contamination of the foreground spectrum by emission from the background object to $< 2\%$.  Using a wider extraction window ($\rm FWHM \sim 1.8\arcsec$) does slightly change the equivalent widths (EWs) of absorption lines (e.g., H9, \ion{Ca}{2}) measured for the foreground galaxy, but it does not affect the results of our stellar population modeling (discussed in \S\ref{sec.fg}).  

We derive redshifts for the two galaxies using IDL code adapted for use in the DEEP2 survey \citep[S. Faber et al. 2009, in preparation;][]{Coil2004} from the publicly available programs developed for the SDSS.\footnote{$\rm http://spectro.princeton.edu/idlspec2d\_install.html$}  In brief, this code calculates the minimum $\chi^2$ value 
as a function of the lag between an observed spectrum and a linear combination of three templates.  One template is an artificial emission line spectrum with lines broadened for consistency with an instrumental resolution of $\rm FWHM \sim 235~\rm km~s^{-1}$, one is the coadded spectrum of thousands of absorption-line dominated galaxies \citep{Eisenstein2003}, and one is an A star spectrum.  The spectral region which includes interstellar UV absorption lines was not used to constrain the fits.  Each galaxy redshift has also been confirmed by eye.  We find that the redshifts of the background and foreground galaxies are $z=0.694248 \pm 0.000001$ and $z=0.47285 \pm 0.00002$, respectively.  The errors are the formal $1\sigma$ uncertainties in the fitted values.  These redshifts are within $\lesssim 92~\rm km~s^{-1}$ of the redshifts obtained in the Team Keck Treasury Redshift Survey \citep[TKRS;][]{Wirth2004} for these objects.  We define the systemic velocity of each galaxy to be at the corresponding fitted redshift.

\begin{figure*}[ht]
\begin{center}
\includegraphics[width=3.in,angle=90]{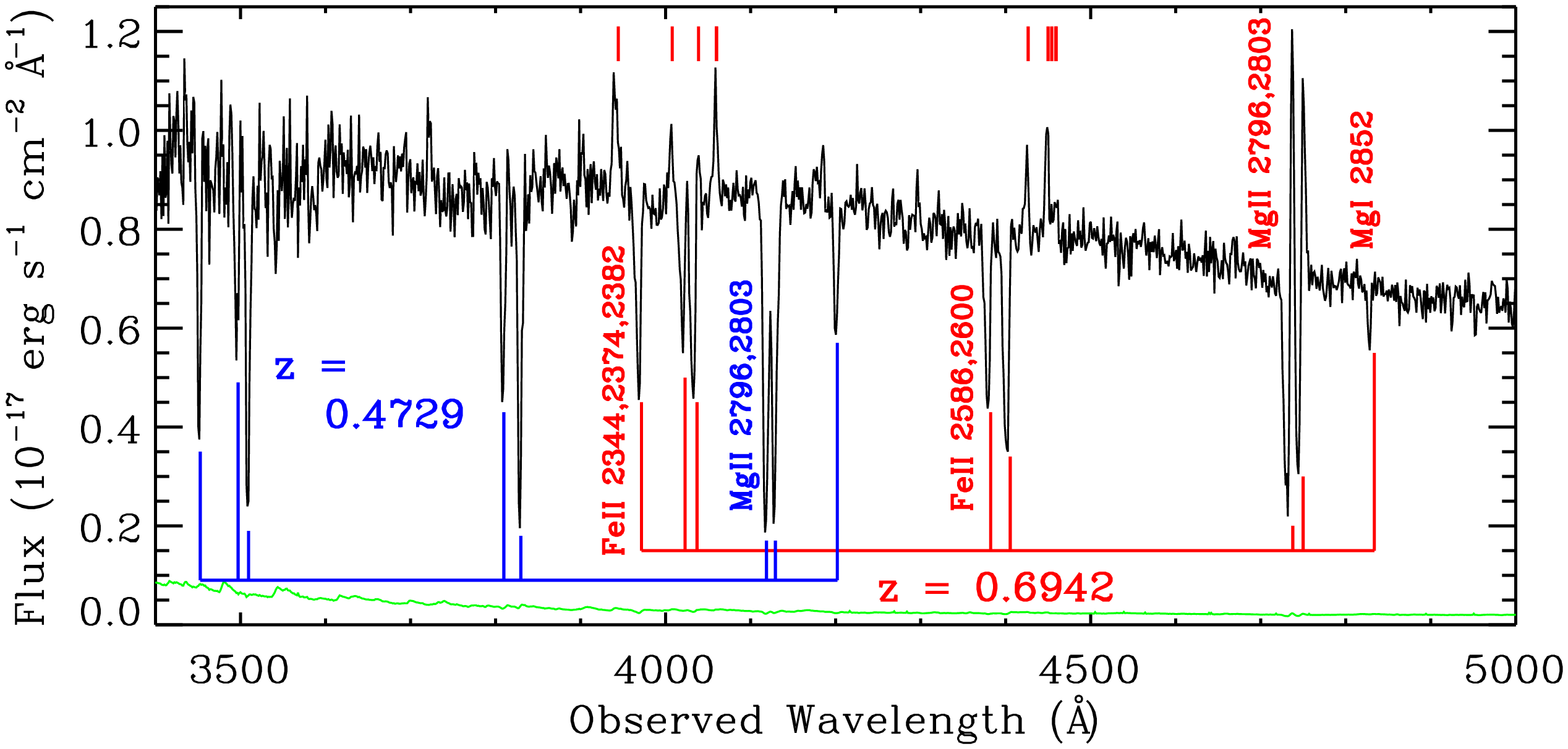}
\caption{Keck/LRIS spectrum (unsmoothed) obtained with our slitlet observations (with orientation shown with the solid lines in Figure~\ref{fig.im}) of the background galaxy. This galaxy exhibits \ion{Mg}{2}, \ion{Mg}{1} and \ion{Fe}{2} absorption offset by $\sim \rm -300~ km~s^{-1}$ from systemic velocity (shown in red), indicating an outflow of cool gas.  \ion{Fe}{2} fine-structure transitions at the systemic velocity of the background galaxy are marked above the spectrum with short vertical red lines.  The green line at  flux levels $\approx 0.1\sci{-17} \rm erg~s^{-1}~cm^{-2}~\AA^{-1}$ shows the $1\sigma$ error in each pixel.  
Strong \ion{Mg}{2}, \ion{Mg}{1} and \ion{Fe}{2} absorption features from the foreground 
galaxy are shown in blue. 
\label{fig.fullspec}}
\end{center}
\end{figure*}

Figures~\ref{fig.im} and~\ref{fig.fullspec} show HST imaging of the galaxy pair and our  LRIS spectrum of the background galaxy, respectively.  UV absorption lines in the rest-frame of the background object are marked in red and are offset to negative velocity from systemic, indicating the presence of an outflow.  
Emission due to \ion{Mg}{2}, as well as emission near the \ion{Fe}{2} absorption lines, is also evident.
The same lines in absorption are marked at the systemic velocity of the foreground galaxy in blue; these marks coincide with strong absorption in the spectrum, and are presumably due to gas associated with 
the $z=0.4729$ foreground galaxy.  

We measured EWs of absorption and emission features after normalizing the spectra to the continuum level.  This level was determined via a linear fit to the continuum around each feature of interest.  We selected different continuum regions for each transition with widths between 8 \AA\ and 72 \AA.
We used a feature-finding code described in \citet{Cooksey2008} to identify and measure the boxcar EW in both emission and absorption lines with $\rm EW / \sigma_{EW} \ge 3$ 
(Table~\ref{tab.ew}).
The foreground galaxy halo/disk absorption has rest equivalent width $W_r (2796) = 3.93 \pm 0.08$ \AA \ and is classified as an ``ultrastrong" \ion{Mg}{2} absorber \citep{Nestor2007}.  $W_r (2796)$ values this large are rare in surveys of \ion{Mg}{2} absorption in QSO 
sightlines; e.g., absorbers with $W_r (2796) > 3.5$ \AA \ make up only 3\% of the population of all absorbers with $W_r(2796) > 1$ \AA \ \citep{Prochter2006}.  Although the number of known galaxy-QSO absorber pairs with impact parameters less than $ 20~\rm kpc$ is quite small \citep[$\sim 30$ from][]  {Bouche2007,Kacprzak2007,Churchill2005intermediatez,Steidel1995}, only two of these have $W_r (2796) > 3.5$ \AA.  (We are not including the  galaxy-absorber pairs from the sample of \citet{Nestor2007} here, as they lack spectroscopic confirmation of the redshifts of the associated galaxies.)  This system also exhibits $W_r (3935) \sim 1$ \AA \ and is classified as a ``strong" ($W_r (3935) > 0.5$ \AA) \ion{Ca}{2} absorber  \citep{Wild2006}.  The $W_r (3935)$ of this halo/disk gas is among the largest measured for the absorbers in the \citet{Wild2006} study.

We find no evidence of emission (e.g., [\ion{O}{2}] $\lambda \lambda 3727, 3729$) due to another galaxy at the redshift of the foreground absorber in either the slitmask or the longslit spectrum.  Furthermore, the other objects identified in the HST image are significantly fainter and lie at larger impact parameters than the previously identified foreground galaxy.  We conclude that the foreground absorption is associated with this galaxy located at $\rho = 16.5~ h_{70}^{-1} \rm~kpc$ from the center of the background galaxy, where the center is determined from quantitative morphological analysis of the $i_{775}$-band image by J. Lotz (2008, private communication), using the method described in \citet{Lotz2004,Lotz2006}.

\section{ANALYSIS OF THE GALAXIES}\label{sec.analysis}

\subsection{TKRS4389 (Background Galaxy)}\label{sec.bg}
\begin{figure}[ht]
\begin{center}
\includegraphics[width=1.8in,angle=90]{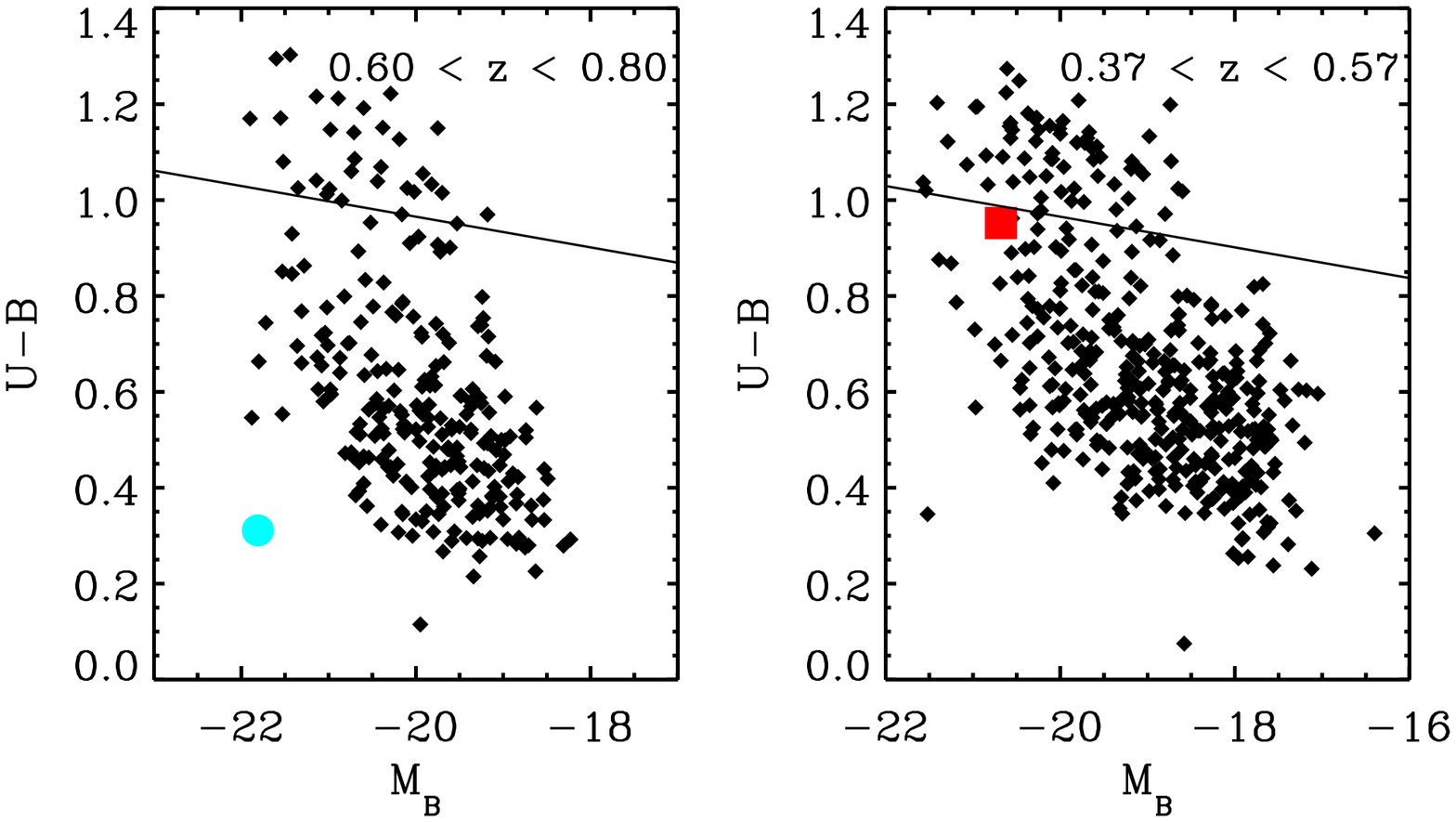}
\caption{$U-B$ vs.\ $M_B$ for TKRS galaxies with $0.60 < z < 0.80$ (left-hand panel) and $0.37 < z < 0.57$ (right-hand panel).  Background (TKRS4389) and foreground (TKRS4259) galaxies are marked with a large circle and square, respectively.  The solid line divides the ``red sequence" from the ``blue cloud" as given in \protect \citet{Willmer2006}.  The foreground galaxy lies in the so-called ``green valley" and may evolve to the bright end of the red sequence \protect \citep[e.g.,][]{Faber2007}. \label{fig.cmd}} 
\end{center}
\end{figure}

The background galaxy is one of the brightest galaxies in the $B_{435}$-band 
in the TKRS \citep{Wirth2004} at its redshift 
and is among the bluest objects in the ``blue cloud" in the galaxy 
color-magnitude diagram (CMD), as shown in the left panel of Figure~\ref{fig.cmd}.  
We derive a star formation rate (SFR) of $\sim 80\rm~M_{\odot}~yr^{-1}$ using 
the luminosity of the H$\beta$ line measured in \citet{Weiner2007} and 
assuming a \citet{CharlotFall2000} dust attenuation curve with effective V-band optical depth $\tau_V \sim 1.5$
as derived from the stellar population modeling discussed below.  We assume case B recombination to 
calculate an H$\alpha$ luminosity and apply the calibration of \citet{Kennicutt1998} to convert this to SFR.  
\citet{KK2004} measure an oxygen abundance of $\rm 12 + log(O/H) = 8.76 \pm 0.15$, close to 
the solar abundance value of 8.72.
Our LRIS spectrum reveals a weak [\ion{Ne}{5}] $\lambda 3426$ emission line, which indicates the 
object is host to an AGN \citep[e.g.,][]{Ho2008}.  AGN activity may therefore affect the SFR estimate and the estimate of the oxygen abundance.  Because the [\ion{Ne}{5}] emission is weak, a Gaussian fit to the line profile results in a rest-frame dispersion of $125 \pm 40~\rm km~s^{-1}$; this value is within $\sim 0.8\sigma$ of the instrumental velocity dispersion at this wavelength.  (All other velocity dispersions reported are calculated by performing a nonlinear least-squares fit of a Gaussian to the appropriate emission line and its surrounding continuum and subtracting the instrumental velocity dispersion from the fitted Gaussian $\sigma$ in quadrature.  Continuum regions were chosen to extend at least 7 \AA \ from line center.)  We also observe strong \ion{Mg}{2} emission 
that results in a P-Cygni line-profile (Figure~\ref{fig.fullspec}), as well as  
a series of emission features near the \ion{Fe}{2}
resonance lines which we identify as fine-structure transitions from excited states of \ion{Fe}{2}.  To our knowledge, this is the first reported case of
\emph{narrow} fine-structure \ion{Fe}{2} emission from any extragalactic object. 
These emission features may be related to AGN activity, but perhaps could be attributed to
some other physical mechanism.  Their origin will be discussed in a future work.  

To estimate the contribution of the AGN to the continuum emission,
we used \citet{BC2003} stellar population synthesis models to generate a grid 
of synthetic starburst galaxy spectra.  All models include an old stellar population, specifically a 7 Gyr-old single-burst stellar population (SSP7) with solar metallicity, and  
the stellar population of an 100 Myr-old ongoing burst with constant star formation rate. 
These particular choices of old and bursting stellar populations are motivated by \citet{Yan2006}, who use them to identify post-starburst galaxies via stellar population modeling.

To each of these spectra, we add
a featureless power-law continuum with a variable normalization to model
the AGN contribution to the galaxy continuum.  We also allow for a variable amount of dust attenuation, parametrized by $\tau_V$.  (Details of the fitting are discussed in \S\ref{sec.fg} and K. Rubin et al. 2009, in preparation.)   
The spectrum is fit well by models with a range in $\tau_V \sim 1.5 - 1.7$ that combine 
a strong starburst
with a power-law component that 
contributes less than 30\% of the emission at 4200 \AA \ in the rest frame.
We conclude that the continuum emission is consistent with a galaxy spectrum that is dominated by intense star formation, rather than
AGN activity.  

Because of its high surface brightness and strong continuum 
blueward of observed wavelength $\rm \lambda_{obs} \sim 5000$ \AA, 
this galaxy is an excellent and rare candidate for 
probing gas in foreground halos.  At the same time, its spectrum
may be used to probe cool gas (via low-ionization absorption) associated with its own ISM and halo but foreground
to its bright star-forming regions.  As noted previously (see \S\ref{sec.thepair}), 
the spectrum of TKRS4389 exhibits evidence for a substantial outflow of cool gas.  
A quantitative analysis of the outflow properties (e.g., velocity, $C_f$, 
limits on the outflow column density, and mass outflow rate) of a large sample of galaxies including this one will be
undertaken by K. Rubin et al. (2009, in preparation).  

\begin{figure}
\includegraphics[width=2.in,angle=90]{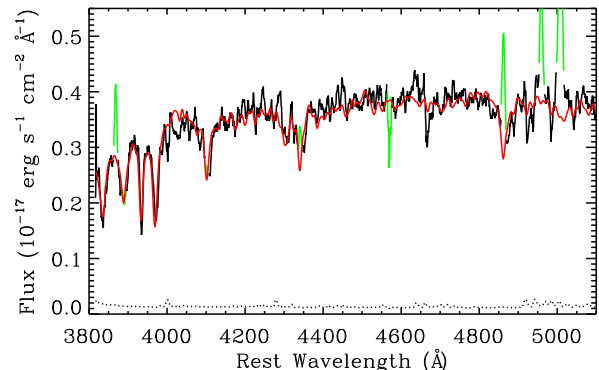}
\caption{Spectrum of the foreground galaxy smoothed over 7 pixels (black) 
and the smoothed post-starburst model spectrum which yields the best fit (with $A = 4$ 
and $t_{burst} = 1.0\rm~Gyr$; red).  The smoothed error values in the data are shown with the dotted line.
Emission lines and the central regions of Balmer absorption lines 
in the galaxy spectrum were masked out prior 
to fitting (masked regions are shown in green), as the model
spectra do not include line emission from AGN or \ion{H}{2} regions.   Bad pixels are also masked out.  Models with 600\,Myr $\leq t_{burst} \leq$ 1.4\,Gyr successfully reproduce the strong Balmer, \ion{Ca}{2} K and G band absorption in the data.  \label{fig.specmodel}}
\end{figure}

\begin{figure}[ht]
\begin{center}
\includegraphics[width=4.in,angle=90]{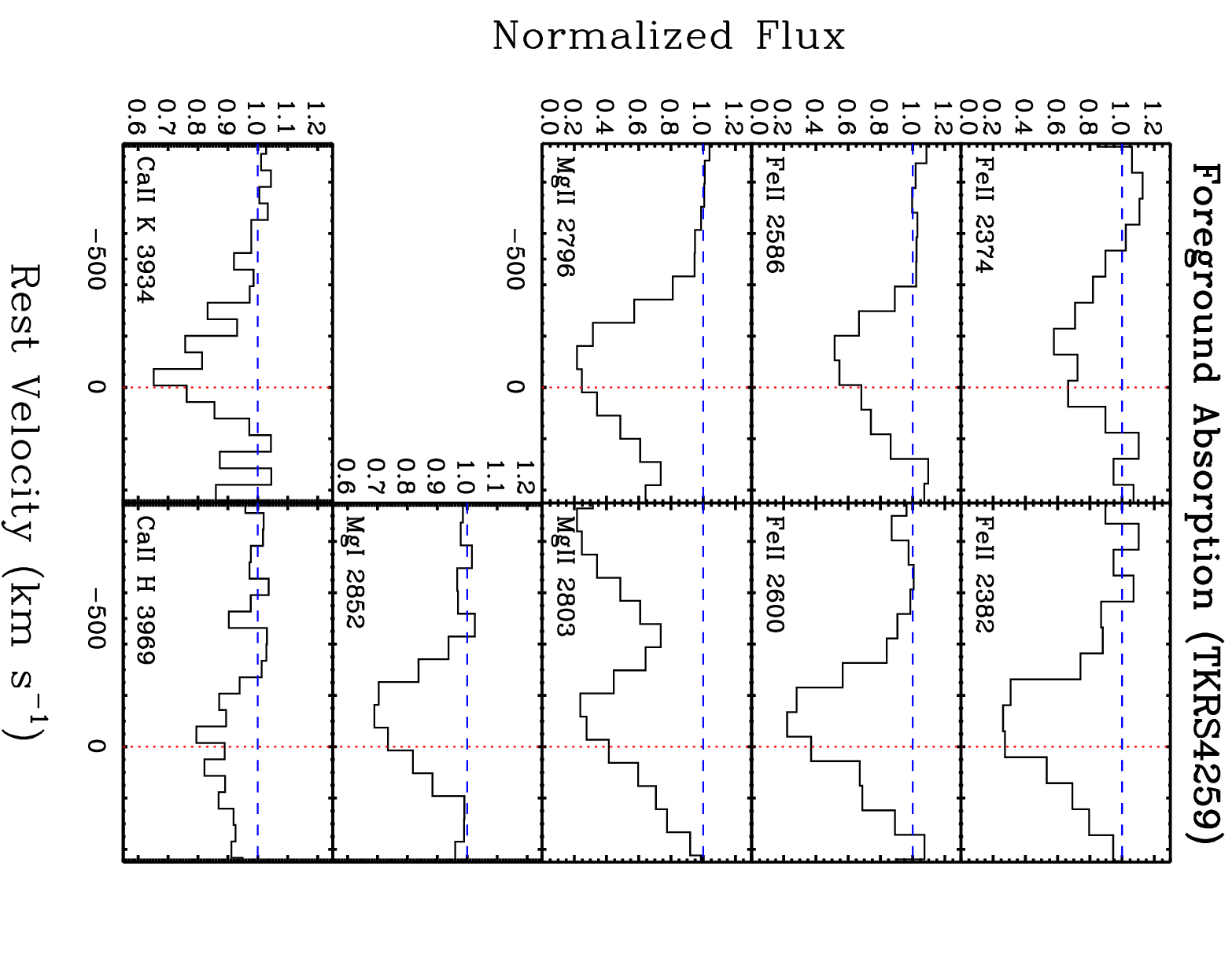}
\caption{UV absorption lines due to the foreground halo in the spectrum of the background galaxy at a distance of $16.5~ h_{70}^{-1}$ kpc.  The systemic velocity is marked with vertical dotted lines.  The dashed line marks the continuum level.  The rest EW of \ion{Mg}{2} $\lambda 2796$ absorption due to the foreground halo is $3.93 \pm 0.08$ \AA.  We note that these absorption profiles are asymmetric about the deepest part of each line.  The \ion{Mg}{2} lines are heavily saturated (see \S\ref{sec.psb_scenario}) and have depths of $\sim 20$\% of the continuum level.  This is indicative of a $C_f$ for \ion{Mg}{2} absorbing clouds of $\gtrsim 0.8$.  \label{fig.fgspecstamps}} 
\end{center}
\end{figure}

\subsection{TKRS4259 (Foreground Galaxy)}\label{sec.fg}
The proximity of this galaxy pair allows us to analyze the halo properties
of the foreground galaxy at an impact parameter of $16.5~ h_{70}^{-1}$ kpc.  To
understand these properties in context, we
first present an analysis of the luminous components of the foreground galaxy.

TKRS4259 is a massive ($\rm \log M_*/M_{\odot} = 11.15 \pm 0.08$, where $\rm M_*$ is the stellar mass; \citealt{Bundy2005}), relatively luminous galaxy, located in the ``green valley" between the red sequence and the blue cloud in the $U-B$ vs.\ $M_B$ CMD \citep[Figure~\ref{fig.cmd};][]{Willmer2006}.  
Although a Gini/M20 analysis (J. Lotz, 2008, private communication) 
indicates TKRS4259  is an early-type galaxy, it shows signs of a 
disturbed morphology or dust lane in 
all of the HST/ACS bands
(i.e., a low surface brightness tidal feature; Figure~\ref{fig.im}).  
Both its location in the CMD and its morphology indicate the galaxy 
may be transitioning from a previously star-forming galaxy to a fully quenched object.  
TKRS galaxies of similar $M_B$ and redshift have approximately solar
oxygen abundance or greater \citep{KK2004}.

The [\ion{O}{2}], $\rm H\beta$, and [\ion{O}{3}] emission-line luminosities 
for this galaxy were measured by \citet{Weiner2007} and are included in Table~\ref{tab.photinfo}.  
We also detect narrow [\ion{Ne}{5}] $\lambda 3426$ emission with rest-frame velocity dispersion $143 \pm 55\rm~km~s^{-1}$ 
(see Table~\ref{tab.ew}), indicating this galaxy hosts an AGN. 
Before calculating a SFR based on the $\rm H\beta$ or [\ion{O}{2}] luminosities, we investigate the possible AGN contribution to this lower ionization line emission.  The line luminosity ratio [\ion{O}{3}]$\rm /H\beta \sim 2.3$; the galaxy falls just below the [\ion{O}{3}]$\rm /H\beta = 3.0$ ratio dividing LINERs (at lower values) and Seyferts (at higher values) in the system of spectral classification described in, e.g., \citet{BPT1981},  \citet{VeilleuxOsterbrock1987} and \citet{Veilleux1995}.  
Unfortunately, we do not have spectral coverage of [\ion{N}{2}] and $\rm H\alpha$ and 
cannot strictly classify this galaxy as a LINER.
However, because [\ion{N}{2}]/H$\alpha$ is correlated with 
metallicity in star-forming galaxies, and because there is a tight relation 
between mass and metallicity in these objects \cite[e.g.,][]{Tremonti2004},
we may use $\rm M_*$ as a proxy for [\ion{N}{2}]/H$\alpha$.  This is demonstrated in Figure 2 of \citet{Weiner2007}, 
which shows a correlation between absolute $H$-band magnitude and both [\ion{N}{2}]/H$\alpha$ and [\ion{O}{3}]/H$\beta$
in blue cloud galaxies. 
Our foreground galaxy lies 
far from the locus of star-forming galaxies in [\ion{O}{3}]/H$\beta$ - $\rm M_*$ space, with a $\rm M_* \gtrsim 20$ times higher than in 
galaxies with similar [\ion{O}{3}]/H$\beta$ ratios in the TKRS.  
It also satisfies the AGN criterion of R. Yan et al. (2009, in preparation), who find that $U-B$ color 
can serve as a proxy for [\ion{N}{2}]/H$\alpha$ as well.
We conclude that LINER or AGN activity dominates the observed line emission.
In addition, \citet{Ptak2007} measure an X-ray luminosity of
$\rm log~L_X = 40.83~(erg~s^{-1})$ for this object.  This luminosity is low enough to be produced by star formation; however, based on our analysis of the optical spectrum, we attribute the X-ray emission to weak AGN activity.

\subsubsection{Ongoing Starburst Scenario}
We cannot, however, rule out ongoing star formation based on this analysis alone.  
The upper limit on the SFR assuming all $\rm H\beta$ emission is due to star formation (and assuming the largest $\tau_V$ from the range derived from stellar population modeling, $\tau_V = 1.8$, as discussed below), 
is $\rm 11~M_{\odot}~yr^{-1}$.  We measure a velocity dispersion of $158 \pm 32~\rm km~s^{-1}$ for this line, and note that a line of this width may arise from an AGN, as narrow emission from AGN has been shown to have approximately the same velocity dispersion as the bulges of the AGN host galaxies \citep{GreeneHo2005}. 
In order to further constrain the rates of current and recent star formation, we  
generate synthetic starburst spectra as described in \S\ref{sec.bg}.   
As above, all models include a 7.0 Gyr-old single-burst stellar population (SSP7) of mass $\rm M_{SSP7}$ with solar metallicity ($Z_{\odot} = 0.02$) and  an 100 Myr-old burst with constant SFR which generates a mass in stars $\rm M_{burst}$.  We generate a similar set of models with supersolar metallicity ($Z = 2.5 Z_{\odot}$). The strength of the burst is parametrized by $A$, where $A = \rm M_{burst} / M_{SSP7}$ \citep[e.g.,][]{Kauffmann2003}.  We create models with $ 10^{-5} < A < 4$ and compare them to the LRIS spectrum of the foreground galaxy at 5620 \AA \ $< \lambda_{obs} <$ 7550 \AA \ (see Figure~\ref{fig.specmodel}).  We do not fit the portion of the spectrum blueward of this range, as it falls on the opposite side of the D560 dichroic, and because it lacks spectral features useful for distinguishing star formation histories.  The portion redward of this range is quite noisy and is subject to atmospheric absorption. 

For each model, we find the best-fit value of $\tau_{V}$, with an attenuation curve parametrized in \citet{CharlotFall2000}, by performing a $\chi^2$ minimization. 
The models do not include nebular or AGN line emission, so emission lines in the data are masked out prior to fitting.  We find that both solar and super-solar metallicity model spectra 
cannot simultaneously match the strengths of the Balmer absorption and the \ion{Ca}{2} K and G band absorption in the data.
The addition of a featureless power-law (QSO) component \citep[$F_{\lambda} = \lambda^{\alpha_{\lambda}}$ with $\alpha_{\lambda} = -1.6$;][]{VandenBerk2001} does not improve the fits significantly.  
These results suggest that there is no ongoing star formation in this galaxy.  A strictly passively evolving model is ruled out by the spectrum as well.

\subsubsection{Post-starburst Scenario}\label{sec.psb_scenario}
We also wish to investigate alternative star formation histories.  One possible star formation history that suggests itself, based on the location of the galaxy in the CMD as well as its morphology, is one in which the galaxy evolved passively for several billion years and then experienced a burst of star formation, which has since ceased \citep[``post-starburst"; e.g., ][]{DresslerGunn1983}.  We again generate a suite of solar metallicity models, all with a SSP which formed 7 Gyr ago.  A burst with constant SFR lasting 100 Myr is added to each model at a time $t_{burst}$, where $t_{burst}$ varies between 1.8 Gyr ago and 200 Myr ago.  As described below, this span in burst age encompasses the range in ages for models which provide a good fit to the data.  Models have values $0.03 < A =\rm M_{burst}/M_{SSP7} < 4$.  We allow for the addition of a QSO power-law continuum component as described above.  We find the best-fit values of $\tau_V$ and the fraction of light contributed by the power-law component by minimizing $\chi^2$ for each model in our grid.   
We perform a visual inspection of the fit of each model, and find that 
models with intermediate $t_{burst}$ values and larger values of $A$ are most successful at fitting the deep Balmer absorption and \ion{Ca}{2} K and G band absorption strengths in the data.
The best-fit model has
a high value of $A$ ($A = 4$; larger values give similar results) and a 1 Gyr-old burst; this model is shown in Figure~\ref{fig.specmodel}.  
The best-fit value of $\tau_V \sim 0.6$, and the fraction of the total light contributed by the power-law component at 4200 \AA \ in the rest-frame is $\sim 15\%$.  Models with $t_{burst} > \rm 1.4~Gyr$, with $t_{burst} = \rm 1.4~Gyr$ and $A < 4$, or with $t_{burst} < \rm 600~Myr$ cannot simultaneously match the Balmer, \ion{Ca}{2} K and G band absorption strengths and thus yield unacceptable fits.
Models with $600~\mathrm{Myr} \leq t_{burst} \leq 1.4~\mathrm{Gyr}$ yield acceptable fits if $A$ is large enough at older ages, with a range in fitted $\tau_V$ values between 0 and 1.8.   
We conclude that a starburst occurred in the galaxy between 1.4\,Gyr and 600\,Myr ago.  
The results are similar for $Z = 2.5 Z_{\odot}$ models.
\\

In summary, we find that the foreground galaxy is host to a low luminosity AGN and has a stellar continuum consistent with that of a post-starburst galaxy.  It has an asymmetric morphology and is located in the ``green valley" in the CMD.  We note that the morphology of this object in the $ V_{606}$ band is in general consistent with that of simulated merger remnants $\gtrsim 0.5~\rm Gyr$ after the final coalescence \citep{Lotz2008}.  These results can be explained by a scenario in which the galaxy recently experienced a merger which triggered a starburst.  The starburst then ceased, perhaps because the cool gas supply was exhausted or expelled. 
In this case, the galaxy will migrate to the red sequence.  
\\

Analyzing our spectrum of the background galaxy 
(TKRS4389) which lies at an impact parameter of $16.5~h_{70}^{-1}~\rm kpc$, 
we identify strong \ion{Fe}{2}, \ion{Mg}{2}, \ion{Mg}{1} and \ion{Ca}{2} absorption from gas associated with 
the foreground galaxy.  A subset of these features is shown 
in Figure~\ref{fig.fgspecstamps} with velocities relative to the systemic velocity of TKRS4259.
The spectrum shown was obtained with the slitlet marked with the solid lines in 
Figure~\ref{fig.im}; however, the spectrum taken with the longslit (marked with dashed lines 
in Figure~\ref{fig.im}) exhibits foreground absorption line profiles with similar depths and 
shapes as those shown.
The measured \ion{Mg}{2}~$\lambda 2796$ EW is extreme:
$W_r (2796) = 3.93 \pm 0.08$ \AA.   
The oscillator strengths of the lines in the \ion{Mg}{2} doublet have a ratio 2:1, and as
the optical depth of the lines increases and they saturate, the EW ratio of the doublet decreases from 2:1 to 
1:1.  The EW ratio of the \ion{Mg}{2} lines in our system is $\sim 1.1$, indicating a high degree of saturation.
In the case of heavily saturated lines, the $W_r (2796)$ value is indicative of extreme kinematics, i.e.,  a
large velocity width of $\Delta v \gtrsim 400~\rm km~s^{-1}$ \citep{Ellison2006}.  
At such a low impact parameter, it is possible that our line of sight 
is probing gas in the outer disk of the galaxy. 
However, even if the galaxy were edge-on (it is not), the predicted differential
rotation of a disk at this impact parameter would be of the order a few tens of $\rm km~s^{-1}$.
We conclude that the gas dynamics are dominated by non-rotational motions
and that the majority of gas is extraplanar, i.e., tracing material in the
halo of TKRS4259. 

Because the \ion{Mg}{2} lines are strongly saturated, the depth of the lines depends 
only on the fraction of the background light source that is covered by absorbing gas ($C_f$) and the 
instrument resolution.  The deepest parts of the line profiles from the spectra taken with slits at \emph{both} orientations 
reach $\sim 20$\% of the continuum level, 
which places a lower limit $C_f \gtrsim 0.8$.  The relatively low spectral resolution smears 
the profiles, and the deepest parts of the lines may in fact appear at much lower normalized flux levels when observed at higher resolution.   
This constraint on $C_f$ is particularly striking given the large beam size provided by the background galaxy:
the distance between the 
two brightest knots in this galaxy is $\sim 4~ h_{70}^{-1}$ kpc at $z = 0.4729$.  This may suggest 
that the absorbing gas complex almost completely covers the background source and has a large
velocity dispersion at all the locations probed.  This is consistent with previous results from 
\citet{Rauch2002}, who report velocity coherence in \ion{Mg}{2} absorbing clouds on $\sim 0.4~h_{70}^{-1}$ kpc scales.  
We leave detailed modeling of the effect of such a large
beam size on the observed absorption to a future work (although see \citealt{Frank2007} for an analysis of the effects of GRB vs.\ QSO beam size on \ion{Mg}{2} absorption system studies).

\section{DISCUSSION}\label{sec.discussion}

Here we examine several possible scenarios for the origin of the \ion{Mg}{2} gas in the foreground galaxy halo.

\subsection{Multiphase Cooling of the Halo / Cold Inflow}\label{sec.multiphase}
In the galaxy formation scenario proposed by \citet{MB2004} and \citet{Mo1996}, $\sim 10^4\rm~K$ clouds condense out of the hot gas in galaxy halos as it cools.  The velocity dispersion of these clouds is approximately that of the galaxy halo; we may therefore estimate the expected velocity dispersion in \ion{Mg}{2}-absorbing clouds given a halo mass for the galaxy.  Using the relation between halo mass and stellar mass derived in \citet{ConroyWechsler2009} via the ``abundance matching" technique, we estimate that the halo mass ($\rm M_h$) of the foreground galaxy is $\rm \log M_h/M_{\odot} \sim 12.9$ for $H_0 = 70~\rm km~s^{-1}~Mpc^{-1}$.  Using the relations between $\rm M_h$ and halo virial velocity given in \citet{MB2004} and assuming a 
singular isothermal sphere for the halo density profile, 
we find that the expected FWHM of the line-of-sight velocity distribution for cool clouds in a halo this massive is 
$\sim 490\rm~km~s^{-1}$.
This dispersion could easily produce a $W_r (2796) > 3$ \AA \ \citep{Ellison2006}, consistent with the measured $W_r (2796)$ for this galaxy.  

The thermal stability analysis of the hot gas surrounding galaxies performed by \citet{Binney2009} suggests that in the case of an isolated halo more massive than that of the Milky Way \citep[$\rm M_h = 10^{12}~ M_{\odot}$;][]{Klypin2002}, a smoothly stratified hot gaseous corona is stable to thermal perturbations and thus will not condense into cool clouds.  However, this study does not account for the effects of gas inflow from the IGM; nor does the assumption of a smooth hot gaseous halo likely apply in this case, given that our host galaxy is a merger remnant.  The three-dimensional cosmological simulations of \citet{Keres2009} show that even in halos well above the ``transition" mass ($\rm M_h \sim 2-3 \times 10^{11}~M_{\odot}$), in which hot, virialized atmospheres develop via shock-heating during gas accretion, cold filaments of gas from the intergalactic medium (IGM) can penetrate deep into the hot halo.  At high redshifts ($z \gtrsim 2$), these cold flows reach the central galaxy and fuel star formation.  At $z \sim 1$, in the case of a $\rm 9 \times 10^{12}~M_{\odot}$ halo, the filaments penetrate to only half the virial radius, but can fragment into dense, cold clouds via shocks, cooling instabilities, or other mechanisms.  As these clouds are   
virialized within the halo, they will acquire a large velocity dispersion and produce a large $W_r(2796)$, as discussed above.

However, we do not consider cold inflow from the IGM to be the most likely origin of the cool halo gas we observe.  First, our detection of $\sim 3.9$ \AA \ \ion{Mg}{2} $\lambda 2796$ halo absorption and $\sim 1$ \AA \ \ion{Ca}{2} K halo absorption in this system (as noted in \S\ref{sec.thepair}; see also Figure~\ref{fig.fgspecstamps} and Table~\ref{tab.ew}) may indicate that the absorbing gas is enriched with metals and dust.  We cannot measure the metallicity or dust content of this system directly, as we lack the necessary spectral coverage.  However,
strong \ion{Mg}{2} absorption systems in the redshift range $0.4 < z < 2.2$ have been shown to significantly redden background QSOs \citep{Menard2008}; furthermore, the degree of reddening increases with $W_r(2796)$.  \citet{Menard2008} find that for absorbers with $W_r(2796) \sim 3.9$ \AA \ at $z=0.4729$, the mean rest-frame $E(B-V) \sim 0.05$ (see their Equation 18).  They also find that absorbers with $W_r(2796) > 1$ \AA \ have an average dust-to-metals ratio close to those measured in the Milky Way ISM.  
Similar results for the mean dust-to-metals ratio of strong \ion{Ca}{2} absorbers have been reported by \citet{Wild2006}, 
who additionally show that dust content increases with larger $W_r (3935)$ \citep[see also][]{WildHewett2005}.  
As noted above, the \ion{Ca}{2} absorption we observe has an $W_r (3935)$ value as large as the strongest absorbers in these \ion{Ca}{2} studies.  These findings suggest that there may be a substantial amount of dust in this system.  
Second, we do not resolve individual gas clouds because of our low spectral resolution ($\rm \sim 290~km~s^{-1}$); however, we note that the absorption line profiles in Figure~\ref{fig.fgspecstamps} exhibit asymmetry.  \citet{Prochaska1997} find that cool gas clouds distributed such that the gas density is proportional to the total matter density in an isothermal spherical halo tend to produce a symmetric velocity profile, although this depends on the number of absorbing clouds in the halo and the particular sightline observed.  
Both the likely high dust content of the absorption system and the line profile asymmetry suggest an alternate origin for the observed cool gas.  

\subsection{Ejection of ISM into the Halo}
We next consider whether the observed halo gas could have plausibly originated in the ISM of the host galaxy or progenitor galaxies which was ejected during a previous starburst, merger or luminous AGN phase.
If this phase occurred $\sim 1$ Gyr ago (as is likely; see \S\ref{sec.psb_scenario}), cool outflows would have had to reach speeds of only $\sim 16~h_{70}^{-1} / \rm cos(\theta)~km~s^{-1}$ 
to reach a projected distance of $16.5~ h_{70}^{-1}$ kpc today, where $\theta$ is the angle between the plane of the sky and the path taken by the gas.  Cool outflows commonly reach speeds in excess of $\sim 100\rm~km~s^{-1}$ in starburst galaxies \citep{Weiner2009,Martin2005,Rupke2005b} 
and have even been observed in post-starburst galaxies at intermediate redshifts \citep{Tremonti2007,Sato2009}.  
Cool gas which has been tidally stripped from the progenitor galaxies of this merger remnant may also give rise to enriched halo absorption with large $W_r (2796)$, provided that the velocity spread in the stripped clouds is large enough \citep{Wang1993}.  The Magellanic Stream, for instance, has a velocity dispersion of $\rm \sim 15-20~km~s^{-1}$ \citep{Bruns2005}, and so could not produce the observed $W_r (2796)$ in the absence of other clouds.
Additionally, anisotropic ejection of gas clouds or tidal streams could easily produce asymmetric absorption line profiles.  We conclude that cool ISM which was driven or stripped out of the galactic star-forming regions during a past starburst, merger or AGN phase is a plausible origin for this gas.  If in reality the gas originated in a cold inflow or halo condensation as discussed in \S\ref{sec.multiphase}, our finding that the galaxy experienced a starburst $\sim 1$ Gyr ago is simply a coincidence, and is not necessarily related to the observed absorbing gas.  We note that it is unlikely that there is ongoing expulsion of ISM, as there is little current star formation activity in the galaxy, and the observed AGN activity 
is weak.  
Therefore, in this scenario, once the cool gas was ejected or stripped, it is likely to have remained cold for approximately 1 Gyr, i.e., since the starburst activity.  The analysis of \citet{Binney2009} suggests that ejected \emph{hot} gas could not have cooled through condensation, although again, the assumptions made in that analysis may not apply in this case.  

Cool gas clouds ejected or stripped from the ISM and moving through a hot medium may be subject to the Kelvin-Helmholtz instability.  \citet{Vietri1997} showed that such disturbances on the surface of a cloud cannot completely disrupt the cloud unless its cooling time is longer than its sound crossing time.  
Following \citet{MB2004}, we assume an isothermal halo with temperature $\rm T_h = 3.3 \times 10^6 K$ and with gas density profile $\rho_g = f_b  V_{max}^2 / 4 \pi G R^2$, where $f_b = 0.17$ is the cosmic baryon fraction \citep{Spergel2003}, and $V_{max}$ is the maximum rotation velocity of the halo (equal to the virial velocity in this case; see \citealt{MB2004} for details).  We assume that the cool clouds are at $\rm T = 10^4 K$ and that they immediately come into pressure equilibrium with 
the surrounding hot gas as soon as they enter the halo.  The sound crossing time for a cloud is given by the ratio of the cloud radius to the cloud sound speed ($\sim 12~\rm km~ s^{-1}$).  To calculate the cooling time of the cloud, we use Equation 33 from \citet{MB2004}.  We find that the sound crossing time exceeds the cooling time by at least one order of magnitude for clouds with masses $5 \times 10^4 - 5 \times 10^6 \rm M_{\odot}$ out to the halo virial radius ($\sim 390$ kpc).  We expect that these clouds will be stable against the Kelvin-Helmholtz instability.  We also note that the clouds at the upper end of this mass range are evaporated by conduction on timescales of $\gtrsim 2$ Gyr, where the evaporation timescale is given by Equation 35 in \citet{MB2004}.  If we assume that the hot gas profile does not change substantially with time, we can conclude that cold clouds driven into the halo from the ISM during a starburst or merger occurring $\sim 1$ Gyr ago should remain cold today, and indeed for at least 1 Gyr more.  

Outflow during the past starburst, merger or luminous AGN event may have been powered by a variety of mechanisms.  The galaxy evolution model advocated by \citet[][and references therein]{Hopkins2008} suggests that in a gas rich major merger gas is driven to the center of the galaxies once they have coalesced, fueling powerful starburst activity and black hole growth and igniting a luminous QSO.  
Feedback from supernovae explosions is well known to drive outflows in starburst galaxies \citep[e.g.,][]{Heckman1990,Rupke2005b} and could have contributed all of the power needed to accelerate the cool ISM to velocities of $\gtrsim 100\mkms$.   
Simulations of major mergers \citep{Cox2004,Cox2006a} show that they can generate a shock-heated gas wind, possibly forcing cool gas into the halo.  In addition, gas from the ISM that was tidally stripped during a merger event may have been left behind at large impact parameters.  Finally, 
a luminous AGN could have contributed to driving cool gas into the galactic halo and to the quenching of star formation.  In the \citet{Hopkins2008} scenario, the phase in which the remaining ISM material is blown away and the QSO is revealed is expected to occur within a few hundred Myr of the merger itself.  The phase in which the QSO has faded and the merger remnant reddens and exhibits a post-starburst stellar continuum occurs  $\sim 800~\rm Myr$ after the merger.  Our results on the star formation history, colors, and morphology of this galaxy fit well into this scenario. 
However, we note that we have the strongest evidence that the galaxy is post-starburst; the occurrence of a quasar phase is purely speculative.  

We also consider whether clouds ejected from the ISM would have a large spread in velocities of $\gtrsim 400~\rm km~ s^{-1}$ 1 Gyr after the starburst event has ceased.  To determine this, we construct a toy model of the behavior of the outflowing gas inside the halo potential.  To start, we follow \citet{Murray2005} and assume that the velocity ($V(r)$) of the outflowing gas as a function of distance from the starburst region ($r$) is given by $ V(r) = 2 \sigma_h \sqrt{(\frac{L}{L_M} -1) \ln \frac{r}{R_0} + V(R_0)^2}$, where $\sigma_h$ is the halo velocity dispersion, $L_M$ is the lower limit on the starburst luminosity needed to expel a large fraction of the galaxy's gas when it is optically thick, and $R_0$ is the initial radius of the outflow.  This equation applies until the terminal velocity is reached, $V_{term } = 3 \sigma_h$.  For $L/L_M = 2$, $R_0 = 1$ kpc and $V(R_0) = 10~\rm km~ s^{-1}$, the gas is driven to radii of $\sim 60$ kpc during a 100 Myr-long starburst.  The results are insensitive to the choice of $V(R_0)$ as long as $V(R_0) \ll \sigma_h$; however, $L/L_M$ and $R_0$ must have values such that the gas does not fall back onto the galaxy in 1 Gyr.  If $R_0 = 1$ kpc, values of $L/L_M \geq 1.4$ are viable; if $R_0$ is doubled, slightly lower values of $L/L_M \gtrsim 1.3$ yield cloud distances and velocities high enough so that they remain in the halo for the necessary amount of time.

We then assume that the starburst ends abruptly, and that the clouds from the wind are now subject only to the gravitational potential of the halo, and neglect the effect of drag forces from the surrounding hot halo gas.  
By integrating the equations of motion for clouds with distances $ \lesssim 60$ kpc and with initial velocities given by the above equation for $V(r)$, we calculate the final positions and velocities after 1 Gyr.  A bipolar geometry for the outflow with an opening angle of $60^{\circ}$ is assumed, consistent with measurements of the opening angle of the outflow cones traced by H$\alpha$ and CO emission in M82 \citep[e.g.,][]{Heckman1990,Walter2002}, such that our line of sight intersects both outflow cones and is parallel to the axis passing through the centers of the cones.  We calculate the line-of-sight velocities for the clouds that are at a projected distance of $\sim 16.5~ h_{70}^{-1}$ kpc after 1 and 2 Gyr.  The velocity spread is well above $400~\rm km~ s^{-1}$ at both time increments.  While this model is quite rudimentary, it indicates that clouds driven out of a galaxy 1 Gyr ago may have the large velocity dispersion that is observed, and that they will maintain this large dispersion for at least another 1 Gyr.

From all of the evidence discussed above, we conclude that the absorption system is most likely gaseous fragments of the ISM from the merger remnant we observe.  This gas may have been ejected by 
a wind driven by a starburst, merger, or AGN, or tidally stripped from the progenitor galaxies.  If indeed these clouds originated in the ISM, this is one of the first measurements of the radial extent of such outflows or tidally stripped gas in the distant Universe and suggests that it reaches far beyond the stellar disk.  
Because the clouds we observe are likely dusty (see \S\ref{sec.multiphase}), this mechanism may be responsible for distributing the diffuse dust detected in galaxy halos to projected separations much larger than the scales of galactic disks \citep{Menard2009}.  

\subsection{The Relationship Between $M_h$ and $W_r (2796)$}
A complementary method which has recently been used to study the host galaxies of \ion{Mg}{2} absorption systems measures the clustering strength of a large sample of absorbers with respect to that of luminous red galaxies (LRGs), to determine the mean halo masses in which absorbers of varying strengths are found.
These studies have shown that there is a weak anticorrelation between $W_r(2796)$ and the bias of \ion{Mg}{2} systems at $z \sim 0.5$. 
This has been interpreted as an anticorrelation with mean absorber halo mass, significant at the 1$\sigma$ level; i.e., absorbers with 1.0 \AA \ $ < W_r (2796)   \le  $ 1.5 \AA \ have mean halo mass  $\mathrm{M_h} \sim 10^{13.2}~ h_{70}^{-1} ~\mathrm{M_{\odot}}$, while $W_r (2796) \ge \rm 1.5 \AA$ absorbers have mean halo mass $\mathrm{M_h} \sim 10^{12.2}~ h_{70}^{-1}~ \mathrm{M_{\odot}}$ (\citealp{Gauthier2009}; see also \citealp{Bouche2006} and \citealp{Lundgren2009}).  
The number of $ W_r (2796) \sim 4$ \AA \ absorbers in these studies is relatively small; 
however, 
the $\rm log~M_h/M_{\odot} \sim 12.9$ we infer for our foreground absorber, while higher than the mean halo mass of strong absorbers quoted above, is well within the uncertainties in the estimates of the mean halo masses for \emph{both} weak and strong absorbers \citep[see Figure 7c,][]{Gauthier2009}.

One explanation proposed by \citet{Bouche2006,Bouche2007} for the origin of strong \ion{Mg}{2} absorbers suggests that cool gas is driven into galaxy halos via starburst-driven winds.   These authors propose that winds are more likely to remove gas from galactic disks in shallower potential wells, and therefore expect strong absorbers to be preferentially found in lower-mass halos.  \citet{Gauthier2009} show, however, that the clustering of strong absorbers is \emph{unbiased} with respect to dark matter halos and thus strong absorption does not occur preferentially at low masses.  Furthermore, the present study provides one example of strong \ion{Mg}{2} absorption which is most likely due to past outflow or merger activity in a high mass ($\rm \log M_h/M_{\odot} = 12.9$) halo.  Our result implies that while galactic winds may be an important source of cool halo gas, they should not necessarily be the dominant contributor in only low mass halos, and indeed \emph{must} not occur preferentially in lower mass halos if they are to fully explain the bias results of \citet{Gauthier2009}.

An alternative picture is inspired by
simulations of, e.g., \citet{DekelBirnboim2006} and \citet{Birnboim2007}, which provide evidence for the ``transition" halo mass, 
above which cold gas is shock heated as it is accreted onto a halo as discussed in \S\ref{sec.multiphase}.  
 \citet{TC2008} find that an anticorrelation between $\rm M_h$ and $W_r(2796)$, as well as the observed frequency distribution of \ion{Mg}{2} absorbers, can be reproduced in a halo occupation model which includes such a transition mass, so that the most massive halos are mostly too hot to produce large $W_r (2796)$ while lower mass halos contain more cool gas and can give rise to high $W_r (2796)$ systems.    
 As shown in \citet{Keres2009}, cold flows can partially penetrate through the hot gas in massive halos at $z \lesssim 1$ and form cold clouds, which may even then fuel further star formation.  While our galaxy is unusual in that it is post-starburst, the existence of an \emph{enriched} $W_r(2796) = 3.93 \pm 0.08$ \AA \ system in a halo with a mass well above the ``transition mass" is consistent with this picture, and if the observed gas originated in an outflow, implies that (1) the effects of star formation on halo gas are observed well into the mass regime in which hot gas accretion dominates the total gas accretion, or (2) there is a spread in transition mass for different galaxy halos.

\section{CONCLUDING REMARKS}\label{sec.conclusions}
Using multiwavelength analysis of the luminous components of a galaxy at $z = 0.4729$ in concert with spectroscopy of a bright, close transverse background galaxy at $z = 0.6942$, we demonstrate that the cool gas traced by \ion{Mg}{2} absorption in the foreground galaxy halo at impact parameter $16.5~h_{70}^{-1}$ kpc most likely originated in the ISM of the host galaxy or its progenitors.  This galaxy has little ongoing star formation and exhibits only weak AGN activity; however, it experienced a starburst $\sim 1$ Gyr ago, which may have been triggered by a merger event.  The cool halo gas could have easily been driven or stripped away from the star-forming regions during this past violent phase, and could have survived in the hot halo since that time.  Our rudimentary toy model for the motions of clouds driven into the halo by a past starburst suggests that these clouds would have a large velocity dispersion $ \gtrsim 400~\rm km~s^{-1}$ today, consistent with our observations.

We emphasize that the 
analysis of the luminous components of this 
galaxy achieves an unprecedented level of detail in the context of studies of distant
absorption-selected 
galaxies with $W_r (2796) \gtrsim 2$ \AA.  Such detail is possible because of the rich dataset available in GOODS, including
the galaxy redshifts from TKRS \citep{Wirth2004} and our LRIS spectroscopy.  
We were able to identify the post-starburst and age-date it using this dataset; previous works could
not necessarily find such a galaxy, and thus present a less complete view of the host galaxies of strong absorbers.  
\citet{Bouche2007} searched for $\rm H\alpha$ emission near $W_r (2796) > 2$ \AA \
absorbers at $z \sim 1$ to investigate a starburst origin for \ion{Mg}{2} absorption;
this method cannot be used to identify host galaxies without $\rm H\alpha$ emission,
such as post-starbursts with weak AGN activity.  \citet{Nestor2007} investigated
environments of $\rm 3~\AA < W_r(2796) < 6~\AA$ absorbers by imaging the corresponding
QSO fields; however, they lacked spectroscopic confirmation of galaxy-absorber associations 
and have not constrained the star formation history of absorber hosts.  

Our study only begins to demonstrate the power of using background galaxies to probe foreground halo gas. 
Several large spectroscopic surveys have recently concluded or are nearing completion that can be used to identify pairs suitable for a targeted search for foreground \ion{Mg}{2} absorption.  Spectroscopic followup with an instrument such as LRIS requires that the background source be bright and blue so that a continuum signal-to-noise ratio of $\sim 15-20$ at $\sim 4200$ \AA \ can be acquired in a reasonable exposure time (i.e., $ B < 22$ in less than 4 hours).  This $\rm S/N$ is needed to achieve a $5\sigma$ EW detection limit of 0.5 \AA \ for a \ion{Mg}{2} absorber at $z \sim 0.5$. 
There are $\sim 40$ galaxy pairs with redshifts measured as part of the DEEP2 survey of the Extended Groth Strip (EGS) with angular separations $< 15\arcsec$ \citep{Davis2003} which meet these requirements.   
Surveys such as COMBO-17 \citep{Wolf2004} and zCOSMOS \citep{Lilly2007} will provide pair candidates in numbers on the same order as DEEP2, as will the forthcoming VIMOS VLT Surveys \citep{LeFevre2005}.  The PRIMUS survey\footnote{http://cmb.as.arizona.edu/$\sim$eisenste/primus/Home.html} \citep{Cool2008} will take galaxy spectra over 10 square degrees and should enable selection of over 1500 galaxy pairs.  
The Pan-STARRS survey\footnote{http://pan-starrs.ifa.hawaii.edu/public/home.html}, which will image 1200 square degrees in 4 filters to $g < 27$, will provide an unprecedented number of photometric redshifts.  Using the average sampling rate and the average redshift success rate of the EGS portion of the DEEP2 survey \citep{Willmer2006} to scale the number of pair candidates to that expected from a photometric survey, and additionally scaling by the 2400-fold increase in sky coverage with Pan-STARRS, we expect that $> 150,000$ 
galaxy pairs will be available for spectroscopic followup.  

The use of galaxies as background probes is complementary to analysis of GRB sightlines, which have a negligible beam size but similarly allow for close analysis of the absorber host \citep{Pollack2009}.  In good seeing conditions and with extended background galaxies, one may probe multiple sightlines through a given foreground halo.  
Size scales smaller than $\sim 4$ kpc at $z = 0.5$ are not currently accessible because of seeing limitations.  However, with adaptive optics and IFUs on 10m-class telescopes and beyond, spatially extended background galaxies will enable study of the morphology and kinematics of halo absorption in concert with detailed analysis of the luminous components of absorber hosts.

\acknowledgements
The authors are grateful for support for this project from NSF grants AST-0808133 and AST-0507483.
J.X.P. acknowledges funding though an NSF CAREER grant (AST-0548180).

The authors wish to thank C. Tremonti for freely providing her stellar population modeling code, K. Cooksey for providing her absorption feature-finding code and for a careful reading of the manuscript, and S. Patel for providing IDL code to produce color images.  We thank James Bullock, Hsiao-Wen Chen, Sara Ellison, Jenny Graves, Bill Mathews, Crystal Martin, Brice M{\'e}nard, David Rosario, Jeremy Tinker, Chris Thom and Luke Winstrom for interesting and helpful discussions during the analysis of these results.  

The authors wish to recognize and acknowledge the very significant
cultural role and reverence that the summit of Mauna Kea has always
had within the indigenous Hawaiian community.  We are most fortunate
to have the opportunity to conduct observations from this mountain.

{\it Facilities:} \facility{Keck:I (LRIS)}, \facility{HST (ACS)}.


\clearpage

\begin{deluxetable}{lccc}
\tabletypesize{\footnotesize}
\tablecolumns{3}
\tablecaption{Galaxy properties \label{tab.photinfo}}
\tablewidth{0pt}
\tablehead{\colhead{Property} & \colhead{TKRS4389} & \colhead{TKRS4259}
}
\startdata
z\tablenotemark{a} &   0.69425  &   0.47285  \\
$M_B$\tablenotemark{b} & $-21.81$ & $-20.68$  \\
$U - B$\tablenotemark{b} &  $ 0.31$ &   0.95  \\
$R_{1/2}$\tablenotemark{c} (kpc) &  6.15 &  6.85\\   
$\rm \log M_{*}/M_{\odot}$\tablenotemark{d} & $ 10.31 \pm   0.25$ & $ 11.15 \pm   0.08$ \\
$\rm \log L([OII])$\tablenotemark{e} $\rm (erg~s^{-1})$  &  $42.23 \pm 0.01$  & $40.89 \pm  0.21$ \\
$\rm \log L(H\beta)$\tablenotemark{e} $\rm (erg~s^{-1})$ & $41.83 \pm  0.01 $ &  $40.83 \pm 0.06$ \\
$\rm \log L([OIII])$\tablenotemark{e} $\rm (erg~s^{-1})$ &  $42.66 \pm 0.01$ &  $41.20 \pm 0.03$ \\
$\rm \log  L_X$\tablenotemark{f} $\rm (erg~s^{-1})$ &  $41.27 \pm 0.12$ &  $40.83 \pm 0.13$ \\
$\rm Gini/M20 $\tablenotemark{g} & 0.58/$-1.39$ &   0.56/$-1.87$\\
\enddata
\tablenotetext{a}{From our galaxy template fitting.  Redshifts are within $\lesssim 92~\rm km~s^{-1}$ of the values reported in the TKRS \citep{Wirth2004}.}
\tablenotetext{b}{\citet{Weiner2006}; derived from GOODS ACS and ground-based photometry from \citet{Giavalisco2004} and \citet{Capak2004} in combination with the K-correction procedure described in \citet{Weiner2005} and \citet{Willmer2006}.  The errors in the observed optical magnitudes and colors are 0.05-0.07 mag and 0.07-0.1 mag, respectively.  Uncertainties in the K-correction procedure contribute an additional 0.12 mag $1\sigma$ error in $M_B$ and 0.09 mag $1\sigma$ error in $U-B$. }
\tablenotetext{c}{Half-light radius measured from ACS imaging; \citet{Melbourne2007}.  Radii are accurate to within $< 10$\%.}
\tablenotetext{d}{K-band derived $\rm M_*$; \citet{Bundy2005}}
\tablenotetext{e}{\citet{Weiner2007}}
\tablenotetext{f}{\citet{Ptak2007}}
\tablenotetext{g}{Quantitative morphologies measured from the $V_{606}$-band image from J. Lotz, 2008, private communication}
\end{deluxetable}

\begin{deluxetable}{lccc}
\tabletypesize{\footnotesize}
\tablecolumns{4}
\tablecaption{Rest-frame EW measurements \label{tab.ew}}
\tablewidth{0pt}
\tablehead{\colhead{} & \multicolumn{2}{c}{TKRS4389 (Background)} & \colhead{TKRS4259 (Foreground)}\\
                              & \colhead{$z = 0.69425$} & \colhead{$z = 0.47285$} & \colhead{$z = 0.47285$}\\
        \colhead{Transition} & \colhead{$W_r$ (\AA)} & \colhead{$W_r$ (\AA)} & \colhead{$W_r$ (\AA)}}
\startdata
      & \underline{Self Absorption} & \underline{Halo Absorption} &     \\
              \ion{Fe}{2} 2344 & $ 1.99 \pm  0.08$ & $ 1.96 \pm  0.18$ & \nodata \\
              \ion{Fe}{2} 2374 & $ 1.30 \pm  0.08$ & $ 1.31 \pm  0.17$ & \nodata \\
              \ion{Fe}{2} 2382 & $ 1.80 \pm  0.07$ & $ 3.02 \pm  0.18$ & \nodata \\
              \ion{Fe}{2} 2586 & $ 1.82 \pm  0.07$ & $ 1.69 \pm  0.10$ & \nodata \\
              \ion{Fe}{2} 2600 & $ 2.55 \pm  0.08$ & $ 2.94 \pm  0.12$ & \nodata \\
              \ion{Mg}{2} 2796 & $ 2.71 \pm  0.07$ & $ 3.93 \pm  0.08$ & \nodata \\
              \ion{Mg}{2} 2803 & $ 1.61 \pm  0.06$ & $ 3.49 \pm  0.08$ & \nodata \\
              \ion{Mg}{1} 2852 & $ 0.37 \pm  0.06$ & $ 1.15 \pm  0.07$ & \nodata \\
              \ion{Ca}{2} 3934 & $ 1.03 \pm  0.11$ & $ 0.95 \pm  0.10$ & \nodata \\
              \ion{Ca}{2} 3969 & No Coverage & $ 0.79 \pm  0.11$ & \nodata \\   [0.5ex]
\tableline \\ [-1.5ex]
 & \underline{Emission} &  & \underline{Emission \& Self Absorption} \\
   \ion{C}{2}$\rm ]$ 2324-2329 / \ion{Fe}{2}* 2328 & $ -0.98 \pm  0.09$ & \nodata & No Coverage  \\
                                 \ion{Fe}{2}* 2365 & $ -0.31 \pm  0.06$ & \nodata & No Coverage  \\
                                 \ion{Fe}{2}* 2383 & $ -0.20 \pm  0.06$ & \nodata & No Coverage  \\
                        \ion{Fe}{2}* 2396.1/2396.3 & $ -0.71 \pm  0.08$ & \nodata & No Coverage  \\
                                 \ion{Fe}{2}* 2612 & $ -0.54 \pm  0.07$ & \nodata & No Coverage  \\
                  \ion{Fe}{2}* 2626/2629/2631/2632 & $ -1.13 \pm  0.09$ & \nodata & No Coverage  \\
                                  \ion{Mg}{2} 2796 & $ -1.28 \pm  0.06$ & \nodata & $  6.37 \pm  1.58$ \\
                                  \ion{Mg}{2} 2803 & $ -1.40 \pm  0.07$ & \nodata & $  4.47 \pm  1.23$ \\
                    $\rm [$\ion{Ne}{5}$\rm ]$ 3345 & $ > - 0.29$ & \nodata & $ > - 1.62$ \\
                    $\rm [$\ion{Ne}{5}$\rm ]$ 3426 & $  -0.84 \pm  0.10$ & \nodata & $  -6.50 \pm  0.70$ \\
                     $\rm [$\ion{O}{2}$\rm ]$ 3727 & $ -64.24 \pm  0.21$ & \nodata & $ -23.19 \pm  1.08$ \\
                    $\rm [$\ion{Ne}{3}$\rm ]$ 3869 & $  -5.89 \pm  0.13$ & \nodata & $  -5.10 \pm  0.52$ \\
                               $\rm H\gamma ~4341$ & $  -9.55 \pm  0.17$ & \nodata & $    3.18 \pm  0.38$ \\
                                $\rm H\beta ~4862$ & No Coverage & \nodata & $  -4.19 \pm  0.28$ \\
                     $\rm [$\ion{O}{3}$\rm ]$ 4960 & No Coverage & \nodata & $  -5.08 \pm  0.39$ \\
                     $\rm [$\ion{O}{3}$\rm ]$ 5008 & No Coverage & \nodata & $ -13.81 \pm  0.52$ \\
\enddata
\tablecomments{Rest-frame EW measurements from LRIS spectroscopy.  EWs in the upper portion of the table in the middle two columns are for absorption lines measured in the spectrum of the background galaxy (TKRS4389) at the systemic velocity of the background galaxy (left) and the foreground galaxy (right).  The second column in the lower portion contains emission line EWs for the background galaxy, while the right-most column in the lower portion contains emission and absorption EWs associated with the foreground galaxy and measured from the foreground galaxy spectrum.  \ion{Fe}{2} fine-structure transitions are marked with an asterisk. Errors refer to $1\sigma$ uncertainties, and limits are given at the $3\sigma$ level.  Table entries are marked ``No Coverage" when the spectral coverage needed to determine EWs is lacking, except in the case of \ion{Ca}{2} 3969, in which a strong $\rm H\epsilon$ emission line in the background galaxy interferes with a measurement of the \ion{Ca}{2} absorption strength.}
\end{deluxetable}

\end{document}